\documentclass[12pt]{iopart}
\begin{document}
\title[Particle Mass Changes and Topology]{On Particle Mass Changes and GR:\\
Space-time Topology Causes LHC Leakage}
\author{M.\ Spaans}
\address{Kapteyn Institute, University of Groningen, 9700 AV Groningen, The Netherlands; spaans@astro.rug.nl}
\begin{abstract}
It is argued that a change in particle rest mass must involve a multiply
connected space-time topology.
The LHC can probe these topological effects through the particle leakage
($\approx 18$\%) that it experiences in particle mass changing interactions.
With a probability of $\approx 82$\%, this 4-space leakage causes a downward
shift in the observable Higgs mass, from its physical value
of 131.6 GeV to 125.2 GeV.
\end{abstract}

Mach's principle formed one of the great motivations for Einstein's general
theory of relativity (GR); the inertia of matter locally is determined by the
mass distribution of the universe globally.
The Einstein equation, using the equality of inertial and gravitational mass,
expresses this through the interplay between space-time curvature and
energy-momentum.
In particle physics, the Higgs symmetry breaking[1] mechanism is believed to
introduce the rest masses of elementary particles below the, spatially
constant, electroweak energy scale, $E_H=246$ GeV.
If particles enter a measuring apparatus with an energy well above $E_H$ then
the subsequent interactions can restore the electroweak symmetry locally and
temporarily, and observers should find different rest masses.

However, in order to quantify particle mass changes one needs to measure masses
in a region that is not affected by such changes.
I.e., any quantification of particle mass change can only be defined away from
the local interaction region, using agreed upon mass units.
All processes, be they gravitational or non-gravitational, depend on particle
masses, so some form of remote measurement appears to be necessary for mass
changing processes, unless one is content with an a priori given absolute mass
scale.
Therefore, it is appealing to elevate this mass measurement notion to a strict
physical principle that requires any particle mass changing process to be
intrinsically non-local.
That is, any detectable change in particle masses, when new particles are
created under electroweak symmetry restoration, must be linked to an observed
non-locality in the positions of those particles. This immediately implies that
the global mass distribution of the universe is altered, in some way, during
particle mass changing events.

The magnitude of this non-locality can be quantified through a dimensional
argument.
For isotropic space-time, all observers living now can receive information on a
portion of the universe with a size that is equal to the particle horizon, $R$.
With Planck's constant $h$ and speed of light $c$, the ratio
$D\equiv RE_H/hc\sim 10^{44}$ then constitutes the maximum extent over which
any observer can speak of the locations of particles as well as of rest masses.
Afterall, measurements on scales smaller than $l=hc/E_H$ cause restoration of
the electroweak symmetry.
Since all particle mass changing events are an undeniable consequence of the
global mass distribution of the universe, the value of $D$ cannot be exceeded
during particle mass changing events.
Any duration $t=l/c$, in an inertial frame, for which the electroweak symmetry
is restored by an event, should thus allow observers to detect particle mass
changes globally up to a scale $R$. Therefore, $RE_H/hc\sim L/cT$ generally
holds, in the same inertial frame, for the scale $L$ up to which particle
mass changes are observable locally for a duration $T$.
Even for $T$ much smaller than $t$, the value of $L$ is truly
macroscopic and one is confronted with a problem.
How can the non-local effects of particle mass changes, be they in the form of
quantum fluctuations or macroscopic experiments, be carried across such
distances?

Because $E_H$ is very much smaller than the Planck energy, wormholes are not
likely to be useful in this[2,3], while superluminal motion is not allowed
either under Lorentz covariance.
This suggests topological identifications, which are unrestricted in GR, as the
only large scale option[3,4].
The relative size scale argument above implies that topological identifications
are to be made through time, e.g., a three-torus $T^3$ embedded in 4-space. In
fact, in [3,5,6,7] it is shown that a lattice of three-tori, denoted $L(T^3)$,
is the proper quantum space-time topology in the absence of wormholes. This
multiply connected 4-space facilitates quantum superposition and allows for
space-time paths that are globally distinct, even in a macroscopic sense[7].

Consider then particle rest mass changes in an accelerator experiment.
Within a 4-dimensional spherical region $U$, centered on the experiment and
of linear size $\le L$, the probability density to enjoy a topological
identification through time between the center of $U$ and another spatial
point scales like $1/r^3$, for an observer with a measuring apparatus residing
at a distance $r$ from the center of $U$ and on its spatial boundary.
The probability to find a particle with a changed rest mass outside of a radius
$r$, with $cT<r<L$, is then $p(r)=log(L/r)/log(L/cT)=log(DcT/r)/log(D)$,
only weakly dependent on details that affect $L$ or $r$.

The LHC reaches energies in excess of 7-14 TeV, much larger than $E_H$.
Thus, the LHC can induce particle mass changes and probe space-time
topology effects through the particle leakage that it experiences.
I.e., $L\sim 10^6$ km, even if $T$ is as small as the Planck time.
The latter time scale is appropriate for $L(T^3)$, see[3,5,6,7].
One finds a probability $p(r)\approx 18$\% to encounter particles with
changed rest masses beyond $r\sim 10$ m from the experimental event. An
experimental region of $r\approx 10$ m seems reasonable for an apparatus
like the CMS or ATLAS inside the LHC, and the dependence of $p$ on $r$
is logarithmic anyway.

A subtle point arises for (the detection of) the Higgs boson itself.
The Higgs boson must somehow both change its rest mass and generate its rest
mass, relative to $E_H$. This has to involve some mass {\it difference} that
depends on the global nature of the observer and the observee[7], i.e., on
the value of $p(r)$. Hence, the topological identifications on $L(T^3)$ must
affect the mass of the Higgs boson as much as its spatial appearance in
terms of $r$.

Realize then that the topological leakage identified above is a 4-space
phenomenon. Any mass scale $M_0$ has an associated 4-space volume
$V_0=1/M_0^4$. So this $V_0$ is rescaled to $V=(1-p)^{-1}V_0$ through
$L(T^3)$.
For a {\it physical} global Higgs mass $M_0$, the topologically rescaled
{\it observable} local Higgs mass is therefore $M=V^{-1/4}=(1-p)^{1/4}M_0$.
Obviously, only in the limit of large $r$ (approaching an apparatus size
of $L$), the value of $p(r)$ goes to zero and a single global mass is
measured.
In practice, the value of $r$ is much larger than the Planck scale and much
smaller than $L$.
Hence, a fraction $1-p$ (or $p$) of all produced Higgs bosons are detectable
inside (or outside) $r\sim 10$ m at the local mass value $M$.
This while a fraction $p$ (or $1-p$) of all produced Higgs bosons are
detectable inside (or outside) $r\sim 10$ m at the global mass value $M_0$.

Recently, the LHC consortium presented evidence for a $\sim 125$ GeV scalar
boson[8,9].
This result is lower than the physical Higgs boson mass of $M_0=131.6$ GeV
derived in [5]. However, the global topological effect identified above
rescales $M_0$ to an effective observable value of $M=(1-p)^{1/4}M_0$ with a
probability equal to $1-p\approx 0.82$.
One finds $M\approx 125.2$ GeV for $p(r)\approx 0.18$, in good agreement with
current experimental limits.
Also, it is expected that further LHC scrutiny reveals an asymmetric double
peak signature in the $M-M_0$ mass range, with a $M$-peak to $M_0$-peak
amplitude ratio of $(1-p)/p\approx 4.6$.
If these predictions are confirmed, then a topological description of
space-time[3,5,6,7] and particle rest mass seems appropriate, and one that
follows the spirit of Mach's principle[2,3,5,6,7,10].

{}


\smallskip

\noindent {\bf References}\smallskip
\begin{thebibliography}{}

\bibitem[1]{}Higgs, P., 1964, Rev.\ Lett.\ 12, 132

\bibitem[2]{}Wheeler, J.A., 1959, Phys.\ Rev.\ 97, 511

\bibitem[3]{}Spaans, M., 1997, Nuc.\ Phys.\ B 492

\bibitem[4]{}D\"urr, S., Nonn, T.\ \& Rempe, G., 1998, Nature, 395, 33

\bibitem[5]{}Spaans, M., 1999, arXiv:gr-qc/9901025

\bibitem[6]{}Spaans, M., 2013, Jour.\ of Phys.\ Conf.\ Ser., Vol. 410, id.\ 012149

\bibitem[7]{}Spaans, M., 2013, arXiv:1305.4630

\bibitem[8]{}CMS collaboration: Khachatryan, V. et al., 2012, Phys.\ Lett.\ B 716, 30

\bibitem[9]{}ATLAS collaboration: Abajyan, T. et al., 2012, Phys.\ Lett.\ B 716, 1

\bibitem[10]{}Smolin, L., 2005, arXiv:hep-th/0507235

\end{thebibliography}
\end{document}